\documentclass[12pt,a4paper]{article}
\usepackage[utf8]{inputenc}

\usepackage{natbib}

\usepackage{amsmath}
\usepackage{amsfonts}
\usepackage{amssymb}
\usepackage{graphicx}

\makeatletter
\long\def\@makecaption#1#2{%
  \vskip\abovecaptionskip
  \sbox\@tempboxa{#1. #2}%
  \ifdim \wd\@tempboxa >\hsize
    #1. #2\par
  \else
    \global \@minipagefalse
    \hb@xt@\hsize{\box\@tempboxa\hfil}%
  \fi
  \vskip\belowcaptionskip}
\makeatother

\begin{document}

\thispagestyle{empty} \vskip 1cm

{
\center{ \bf \Large Triggered star formation in giant HI supershells: ionized gas.}

\bigskip

% Authors:
{\large O.V.~Egorov,$^{*}$\footnote{e-mail: egorov@sai.msu.ru}
   T.A.~Lozinskaya $^{*}$\footnote{e-mail: lozinsk36@mail.ru},
   A.V.~Moiseev$^{*, +}$}

% Affiliations:
 \center{ \center ${}^*$\it  Sternberg Astronomical Institute of Lomonosov Moscow State University, \newline{}  Universitetskii pr. 13, Moscow 119992, Russia \newline{}
 ${}^{+}$\it Special Astrophysical Observatory, Russian Academy of Sciences, \newline{}
Nizhnij Arkhyz 369167, Russia}

}

\vspace{20pt}

% Abstract:
{ We considered the regions of triggered star formation inside kpc-sized HI supershells in three dwarf galaxies: IC 1613, IC 2574 and Holmberg II. The ionized and neutral gas morphology and kinematics were  studied based on  our observations with scanning Fabry-Perot interferometer at the SAO RAS 6-m telescope and 21 cm archival data of THINGS and LITTLE THINGS surveys.  The qualitative analysis of the observational data  performed in order to highlight the two questions: why the star formation occurred  very locally in the supershells, and how the ongoing star formation in HI supershells rims influence its evolution?
%We noticed that probably the supershells collision might be a key to star formation triggering.
During the investigation we discovered the phenomenon never before observed in  galaxies IC~2574 and  Holmberg II: we found faint giant (kpc-sized) ionized shells in H$\alpha$ and [SII]6717,6731 lines inside the supergiant HI shells.
}

\bigskip

% Keywords

\section*{Introduction.}

Stellar feedback plays significant role in the regulation of the interstellar medium (ISM) structure and the whole galaxy evolution as well. Stellar winds and supernovae explosions create complexes of shells and supershells of ionized and neutral gas; e.g. \citet{bagetakos11} analysed the HI distribution in 20 nearby galaxies
  and revealed about 1000 cavities in their discs with sizes from 80 pc to 2.6~kpc, expansion velocities from 4 to 64~km/s (in the main 10-20 km/s) and ages from 2 to 150~Myr.

The origin of the largest kpc-sized HI holes and shell-like structures (so-called supergiant shell, SGS) has been the subject of debate for more than two decades. In the standard approach based on the \citet{weaver77} model, HI shells result from the cumulative action of multiple stellar winds and supernovae explosions. However it was recognized long ago \citep[see e.g.][]{tenor88, rhode99, kim99, silich06} that this scenario cannot explain the origin of SGS since stellar cluster remnants detected are inconsistent with the input of mechanical energy required by the standard multiple winds and supernovae model.
Recent studies based on Hubble Space Telescope (HST) observations have found that multiple star formation events over the age of the hole do provide enough energy to drive HI hole formation \citep[see e.g.][]{weisz09a,weisz09b,cannon11a,cannon11b}. A number of SGS exhibit signs of expansion-triggered star formation at their periphery. A special interest consists in a detailed analysis of the interaction of stars and gas in the region of new sites of star formation in the walls of SGS, which can elucidate the process of their evolution.

Dwarf irregular galaxies provide the best environments to study the creation mechanisms of giant HI structures with star-formation episodes in their rims. Because of the slow solid-body rotation and the lack of strong spiral density waves which can destroy the giant shells, they grow to a larger size and live longer compared to such structures in spiral galaxies. The overall gravitational potential  of dwarfs is much smaller than that of spiral galaxies, their HI disc scale height is larger and the gas volume density is lower than in spirals. Therefore the same amount of mechanical energy fed to the ISM of dwarf Irr galaxies creates very large long-lived holes with star formation in the walls triggered by their expansion.

During several years we performed the analysis of the ionized and neutral gas structure and kinematics in nearby irregular galaxies: VII~Zw~403 \citep{zw}, IC~10 \citep{ic10}, IC~1613 \citep{lozinsk03}, IC~2574 \citep{ic2574}, Holmberg~II (Egorov et al. 2016, in preparation). In most of them we had a deal with the star formation ongoing in the rims of the HI supershells. In this work we briefly review the results of our study in the galaxies where the different stages of the SGS evolution are clearly seen: IC~1613, IC~2574 and Holmberg~II.

\section*{Observational data}

The observations in the H$\alpha$ emission line were made at the prime focus of the 6-m telescope of Special Astrophysical Observatory of  Russian Academy of Sciences with SCORPIO \citep{scorpio} or SCORPIO-2 \citep{scorpio2} multi-mode focal reducers. The data obtained in  a scanning Fabry--Perot interferometer (FPI) mode were used for the study of the ionized gas kinematics, while the narrow-band H$\alpha$ direct images were used for analysis of the ionized gas morphology. Neutral gas distribution and kinematics were studied using the data of archival VLA HI 21 cm observations from THINGS \citep{things} and LITTLE THINGS \citep{littlethings} surveys. The detailed description of the observational settings are shown in the corresponding papers for each galaxy discussed there.

\section*{Triggered star formation in SGSs}

We considered three galaxies where  the difference in the evolution stage of SGS is most evident. The aim of our analysis was to highlight two questions: what triggers the star formation  at large scales and how the ongoing star formation influences the evolution of the ``parent'' SGS in a neutral gas? We describe below the most interesting results obtained during the study of each of the selected galaxies and discuss the questions above.

\subsection*{IC~2574}

\begin{figure}
\includegraphics[width=0.5\linewidth]{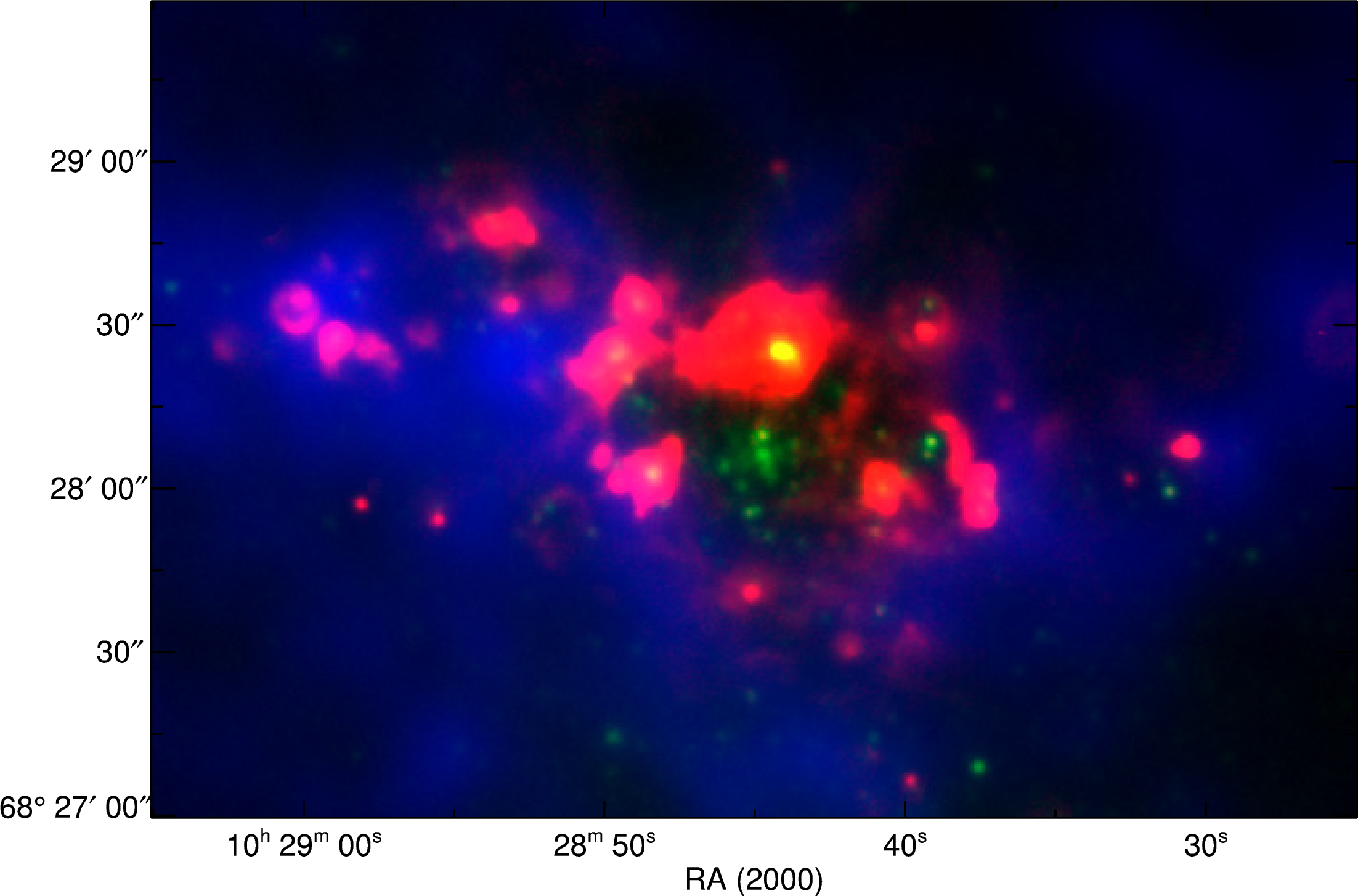}~\includegraphics[width=0.5\linewidth]{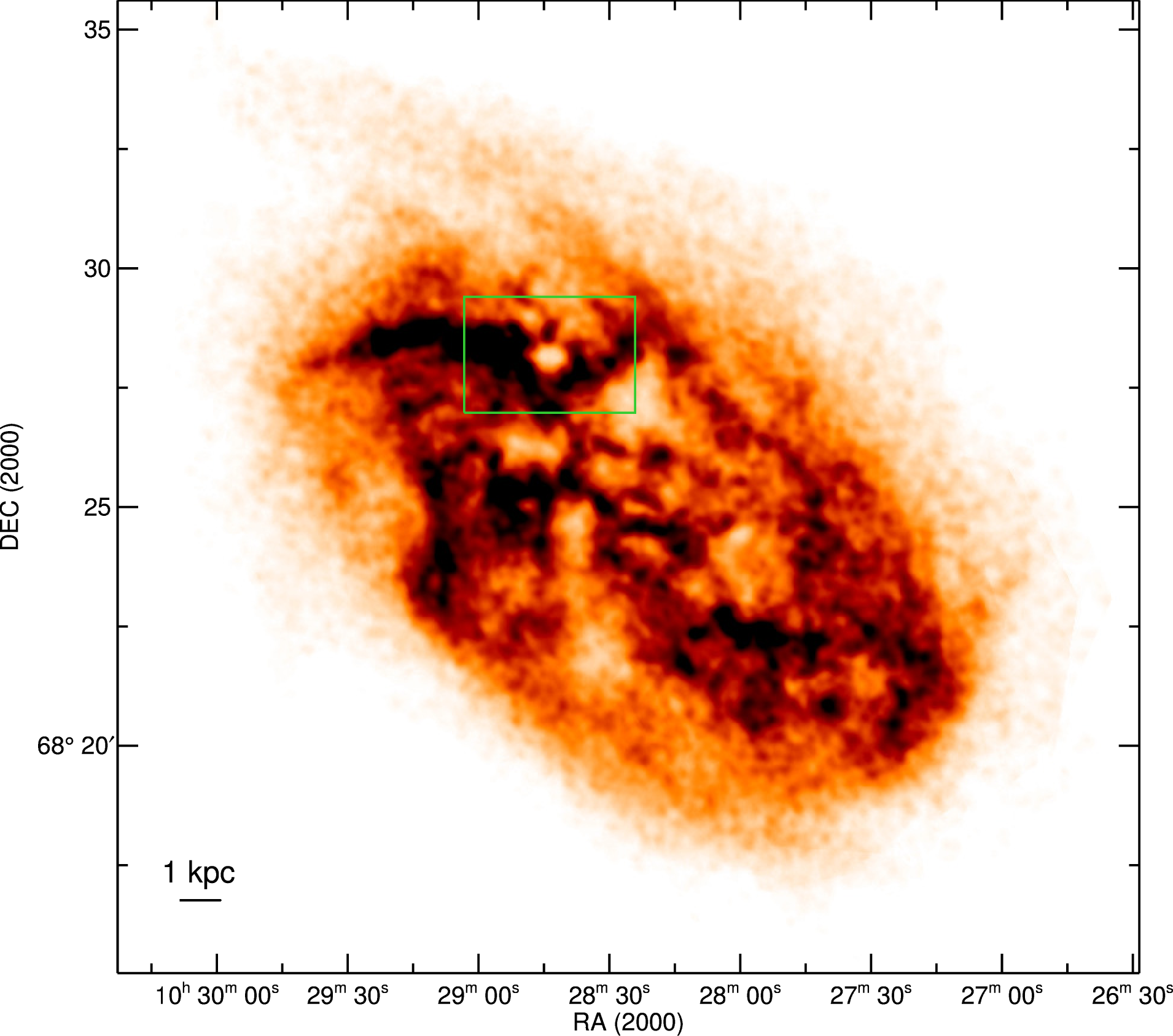}
\caption{IC~2574. Left: False-color composite image of distribution of H$\alpha$ emission (red channel) and stellar continuum (green channel) obtained with 6-m telescope of SAO RAS and HI 21 cm emission (THINGS, blue channel). Right: HI distribution according to THINGS data. Green rectangle denotes the region shown in the left pannel.}\label{fig:ic2574}
\end{figure}

The well known ``supergiant shell'' (\#~35 in the list of \citealt{walter99}) in the dwarf Irr galaxy IC~2574 is located at the north-east outskirts of its disc. This SGS surrounding a $1000\times500$ pc hole in the most prominent region of current and recent
star-formation activity in the galaxy  represents the impressive example  of
giant HI shells with triggered SF along rims (see Fig.~\ref{fig:ic2574}). The analysis of resolved stars based on the HST data \citep{weisz09b} shows that the last most significant
episodes of star formation in SGS began $\simeq$ 100~Myr ago and
the recent bursts of star formation along the walls of the shell are as young
as 10~Myr. The ages of the younger star-formation events are consistent with those derived from broadband photometry and are younger than the estimated kinematic age of the HI SGS, $14 \pm 3$~Myr \citep{walter98, walter99, ic2574}.

The analysis of the ionized and neutral gas kinematics inside the rim of SGS is  presented in \citet{ic2574}. We showed that for almost all HII complexes studied the energy input from young clusters located inside them is sufficient to drive the formation of observed ionized gas structures. The only exception we found is the shell-like region in the north-west part of the SGS that shows a high expansion velocity (65 km/s). This complex is the youngest in the region and its age is 1 Myr.

In the entire region of the SGS in IC~2574 we detected for the first time the faint diffuse emission in both H$\alpha$ and [SII] lines. It seems that the source of the emission observed is the shell-like structure that filled the internal part of the SGS. 
%We made this conclusion based on the ionized gas kinematics analysis performed in \citet{ic2574}.

We conclude  that the observed SGS in IC~2574 is at relatively young stage where the star formation in its walls had been started recently. It is one of the most dynamically active giant HI structures among Irr galaxies. Almost the whole of the rim of the SGS shows the significant increase of star formation rate during the recent 10 Myr. The expansion of the brightest HII region in the north part of SGS rapidly disperses the local HI gas.
One can see here the emission features like ``horns'' located beyond the northern boundary of the SGS.
We can suppose that the destruction of the northern wall of the HI supergiant shell due to star formation will result in the growth of the SGS and in its merging with the neighbouring lower sized HI supershells. After several billion years there will be a system of giant adjoining and/or interacting shell-like HI structures similar to those observed in another galaxy we want to draw attention to below -- Holmberg~II.

\subsection*{IC~1613}

\begin{figure}
\includegraphics[width=0.5\linewidth]{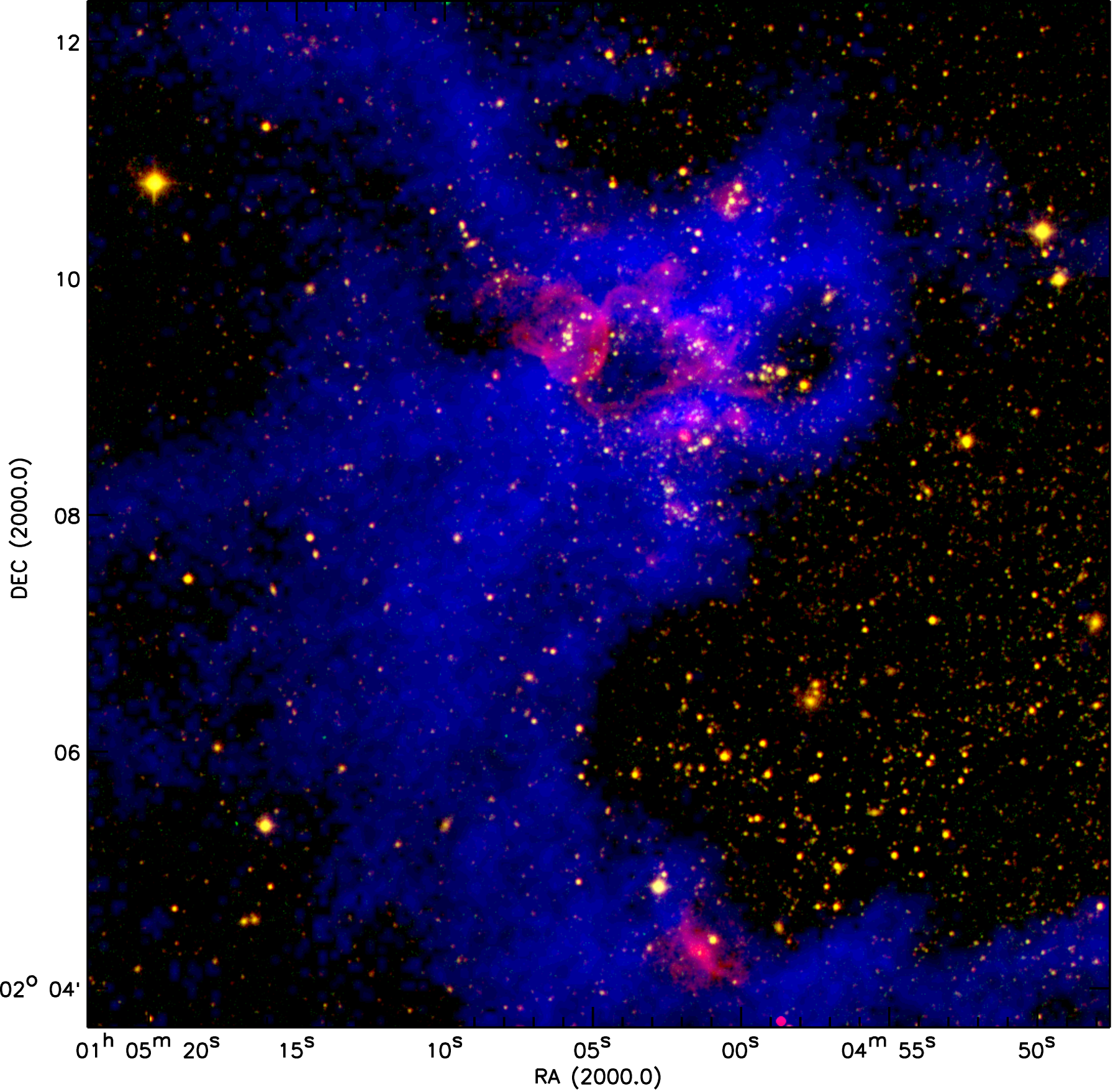}~\includegraphics[width=0.5\linewidth]{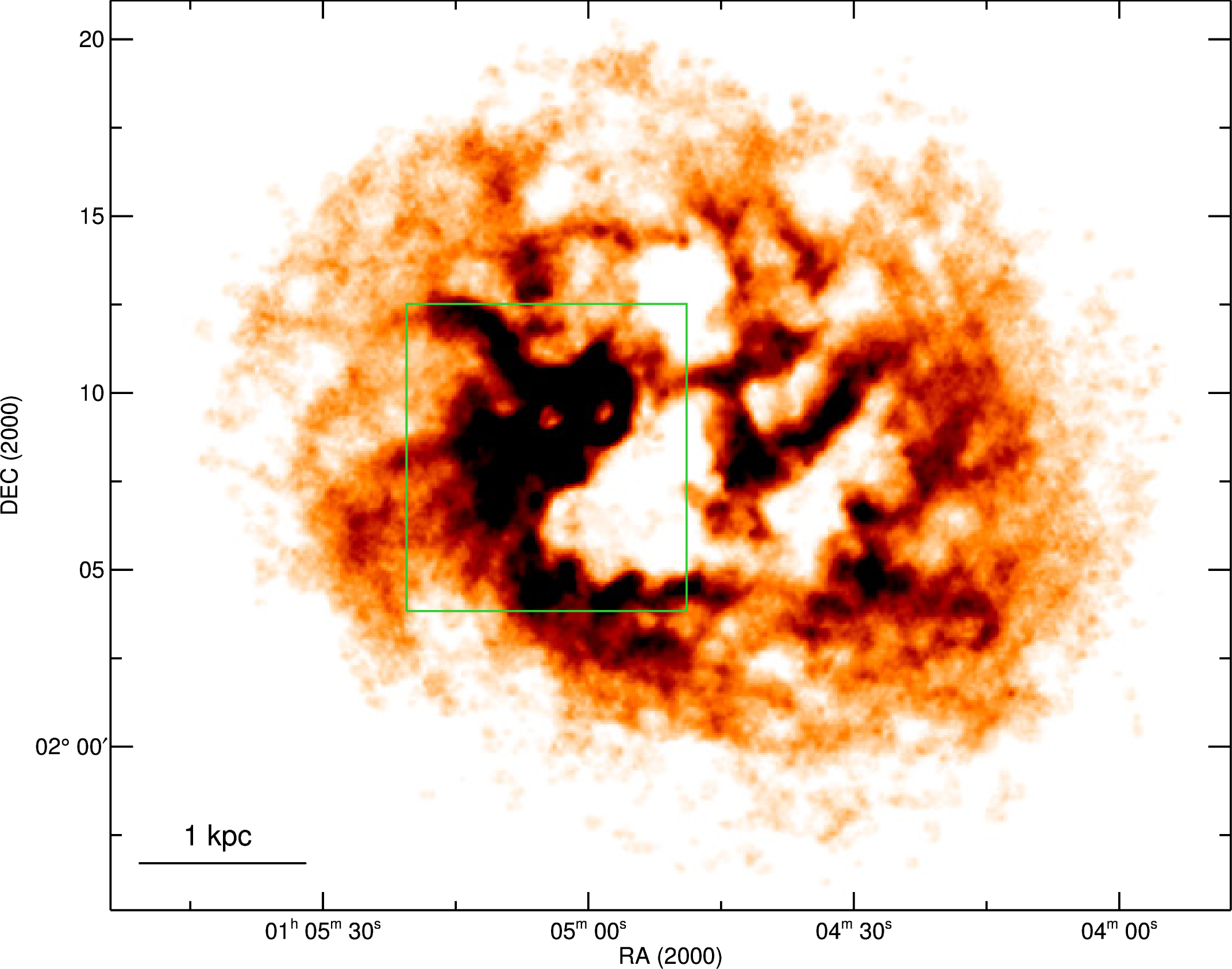}
\caption{IC~1613. Left: False-color composite image of distribution of H$\alpha$ (red color) and stellar continuum (yellow color) obtained with the SAO RAS 1-m telescope  according  to \citet{lozinsketal02}, and HI 21 cm emission (LITTLE THINGS, blue color). Right: HI  distribution obtained by LITTLE THINGS data. Green rectangle  denotes the region shown in the left panel.}\label{fig:ic1613}
\end{figure}

Neutral gas distribution in IC~1613 galaxy shows a highly inhomogeneous structure with a number of HI holes, shells and arc-like structures of different sizes and expansion velocities (see Fig.~\ref{fig:ic1613}). The only known complex of recent star formation is located in the north-eastern rim of the largest (1 to 1.5 kpc sized) HI ``main supershell'' in the galaxy. This supershell is older then SGS in IC~2574 -- its kinematic age no less than 30 Myr \citep{silich06}.
Three extended (300\mbox{--}350~pc) neutral shells with which the brightest ionized shells in the complex of star formation are associated are observed in the direction of the area of highest HI density in the galaxy. Two of H~I
shells were found to expand at a velocity of 15\mbox{--}18~km~s$^{-1}$ \citet{lozinsk03}.
The structure and kinematics of this complex were studied previously in a large number of papers \citep[see, e.g.,][and references therein]{lozinsk03, silich06}.

Fig.~\ref{fig:ic1613} clearly demonstrates that the H$\alpha$ emission in IC~1613 coincides well with the walls of three local HI shells mentioned above. The age of observed H$\alpha$ bubbles are much lower (0.6 -- 2.2 Myr) than the age of the HI shells (5.3 --5.6 Myr), that may be an indirect evidence of the triggering of star formation there by the collision of these neutral gas shells. \citet{lozinsk02} proposed that the whole star formation complex in the rim of the ``main supershell'' is created by the collision of this largest  shell with the giant HI supershell to the north.

\subsection*{Holmberg II}

\begin{figure}
\includegraphics[width=0.5\linewidth]{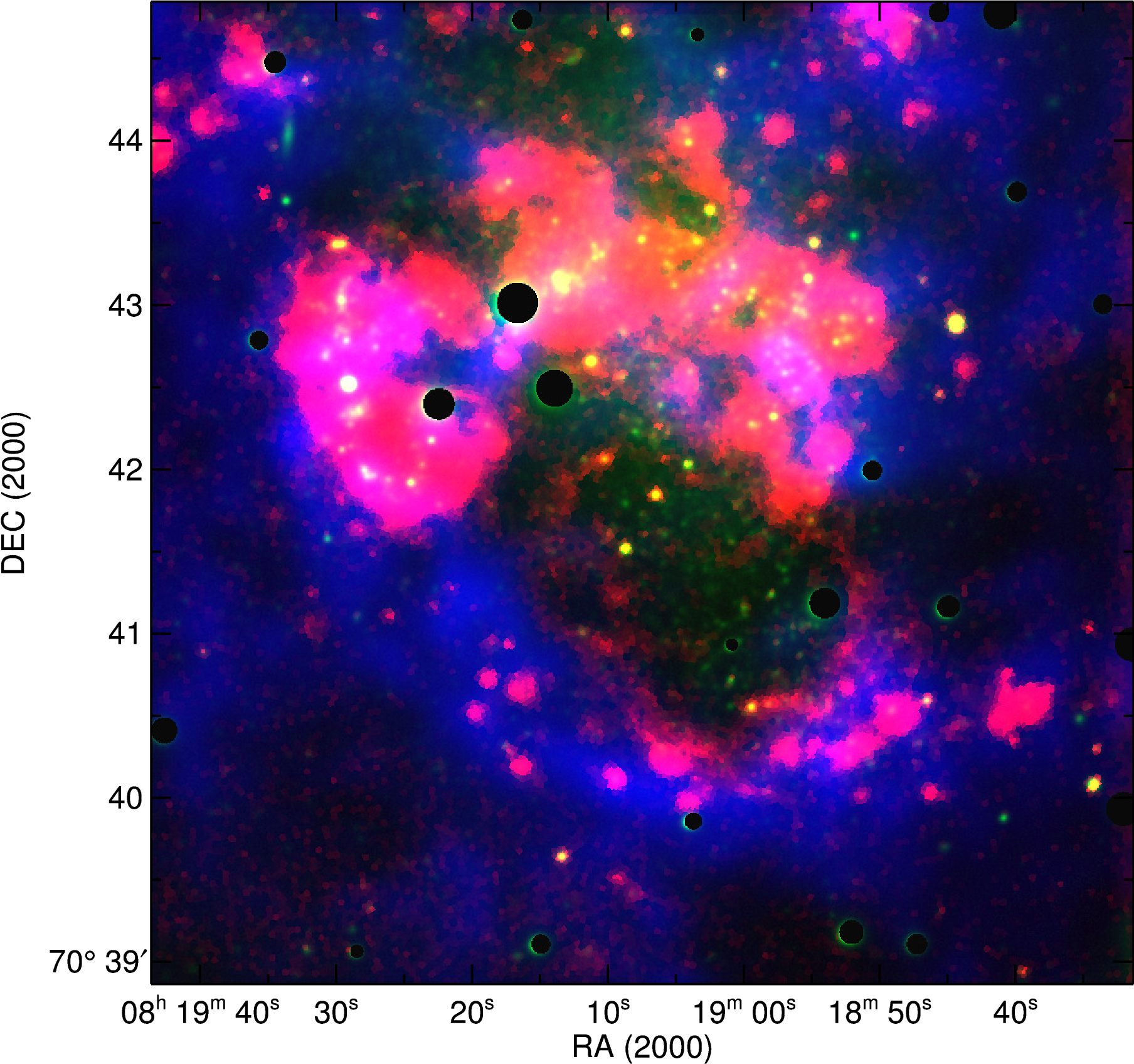}~\includegraphics[width=0.5\linewidth]{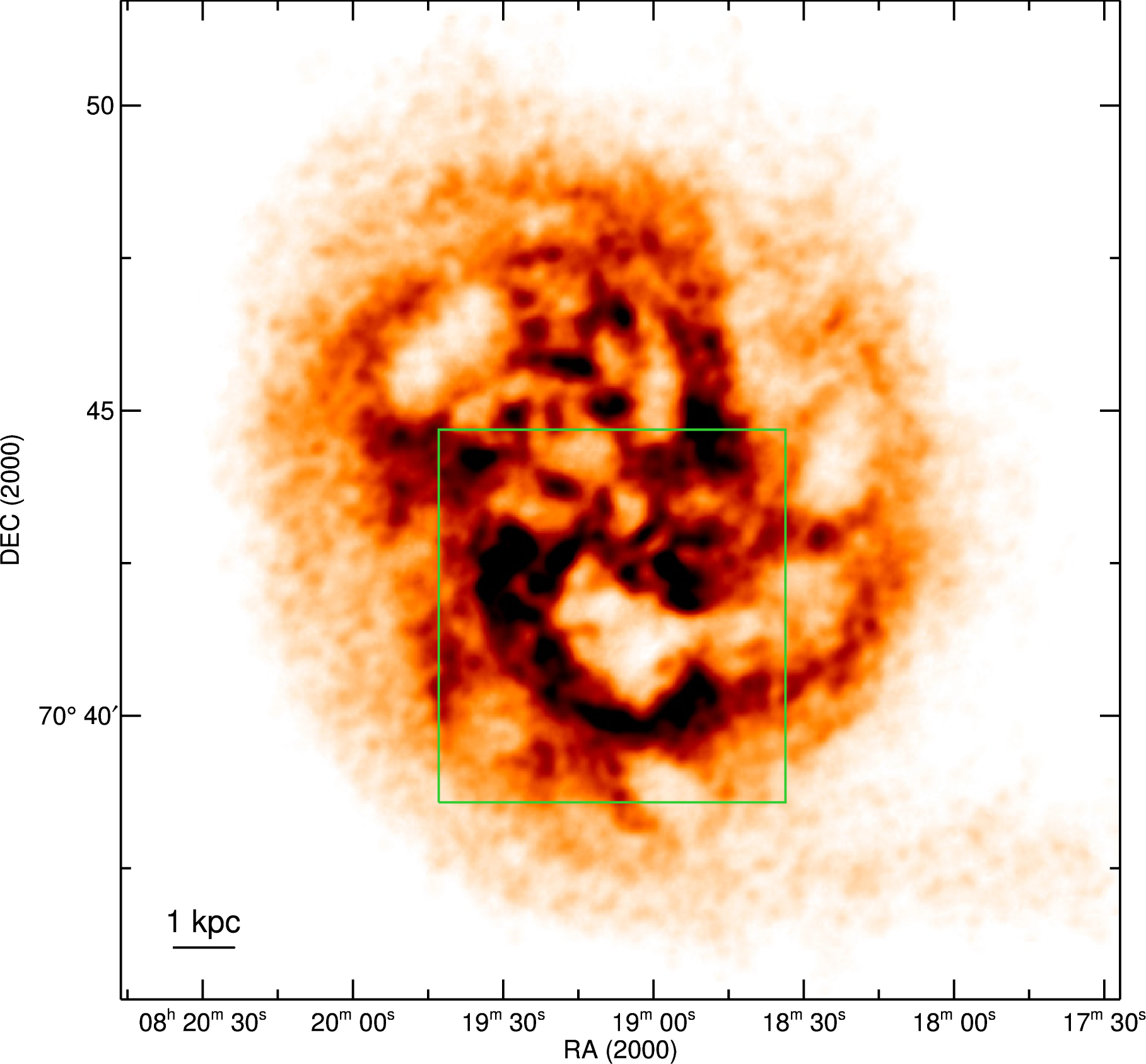}
\caption{Similar to Figure~\ref{fig:ic2574} for Holmberg~II galaxy. 
}\label{fig:hoii}
\end{figure}

Holmberg~II is another example of irregular galaxies that have a non-uniform gas structure with a large amount of shells and holes in the neutral gas distribution (see Fig.~\ref{fig:hoii}). \citet{puche92} found 51 giant holes and supershells in Holmberg~II galaxy; \citet{bagetakos11} revealed 39 neutral gas cavities using more strong criteria. Their sizes are  from 0.26 to 2.11 kpc and expansion velocity -- from 7 to 20 km/s. These values correspond to kinematic age from 10 to 150 Myr. It is necessary to have energy input up to  $10^{53}$ erg for the formation of several SGS.

The object of particular interest is the SGS \#17 \citep[according to the list of][]{bagetakos11} -- the largest and oldest HI supershell in the galaxy. Optical images in the H$\alpha$ line reveal the brightest complexes of star formation located in the rim of this SGS. This supershell is much older than the SGS in IC~2574 described above (150 Myr against 14 Myr). Nevertheless, in both cases we observed that the triggered star formation occurred in the walls of SGS. \citet{weisz09a} analyzed the stellar population using the data of HST observations and showed that star formation occurs in the north rim of the SGS in Holmberg~II about 40 Myr ago and during the last 15 Myr it is observed in the north-east and north-west parts of the shell.

In contrast with the case of SGS in IC~2574, in Holmberg~II we observed that triggered intensive star formation occurred not along the full length of the HI supershell rim, but only in one half of them.
%Performed analysis of the HI distribution showed that the mean HI density as well as mean velocity dispersion are almost not vary along the supershell rim. It is allow us to expect the uniform distribution of triggered star formation complexes along the rim.
Why is the H$\alpha$ emission more intense in the northern part of HI supershell compared with its southern part? Probably, as it was in the case of IC~1613, the interaction and possible collision with two younger and smaller HI supershells to the north from the SGS considered are the reasons for triggering star formation in this region. Indeed, the age of the last starburst episode that occured in the region is in agreement with the estimations of the ages of these two supershells (\#16 and \#22 in the list of \citealt{bagetakos11}) -- 50 and 30 Myr.

It is not surprising to detect H$\alpha$ emission inside the small HI shells in galaxies \citep[see, e.g., LITTLE THINGS survey][]{littlethings}, but whether giant ionized shells could be observed inside the HI SGSs? Up to date there was known only one kpc-sized H$\alpha$ supershell -- LMC 4 in Large Magelanic Cloud \citep{lmc}. Also there were found several supershells with smaller size in that galaxy.
 We already noticed the detection of diffuse ionized kpc-scale supershell in IC~2574 located inside HI SGS, but there we proposed its shell-like morphology only by kinematics analysis. We found the unique structure in Holmberg~II galaxy using the data of FPI and narrow-band imaging observation in H$\alpha$ line. We clearly detected the giant ionized supershell with resolved shell-like structure that has diameter about 2 kpc and coincides with the internal wall of the examined neutral HI SGS in the galaxy. We should note that this structure has a low brightness (about $10^{-17}$ erg/s/cm$^2$/arcsec$^2$). \citet{bastian11} found 126 OB stars in Holmberg~II, seven of them are located inside the H$\alpha$ supershell. These seven OB stars provide enough high energy photons to ionize the internal wall of HI supershell and to create the H$\alpha$ supershell observed. The detailed study of this structure as well as of the gas kinematics in the star formation regions in Holmberg~II will be presented in our forthcoming paper Egorov et al. (2016, in prep.)

The observed parameters of local diffuse ionized gas in the SGS region are similar to those of the
extra-planar diffuse ionized medium (DIG) in spirals and irregular galaxies. The observed  [SII]/H$\alpha$ and [NII]/H$\alpha$ line ratios in the DIG are elevated compared to classical HII regions.
The ionization of DIG in galaxies is traditionally explained by the leakage of ionizing photons from the HII regions as the main source \citep[see, e.g.,][and references in these papers]{seon09, hidgam06}. Because of this similarity we may propose that faint ionized gas emission observed in the SGSs in IC~2574 and Holmberg~II galaxies is the result of leakage from the bright HII regions in the SGSs walls. Indeed, all the HII complexes in the walls of studied galaxies have nonuniform, clumpy, or filamentary structure, which allows radiation to leak outside through low-density regions.

%To explain the increase of [NII]/H$\alpha$, [SII]/H$\alpha$ and [OIII]/H$\beta$ with galactic height, additional %sources of ionization were considered: ionization by shocks or turbulent mixing layers may be responsible for some of %the DIG emission.

%Because of this similarity
% we may  propose another explanation of the faint diffuse ionized gas emission observed in the SGSs in IC~2574 and Holmberg~II galaxies:  the result of leakage from the bright HII regions of both the ionizing photons and mechanical energy of stellar winds and supernovae.
% Indeed, all the HII complexes in the walls of studied galaxies have nonuniform, clumpy, or filamentary structure, which allows radiation and shocks to leak outside through low-density regions.

\section*{Conclusions}

\begin{figure}
\includegraphics[width=0.5\linewidth]{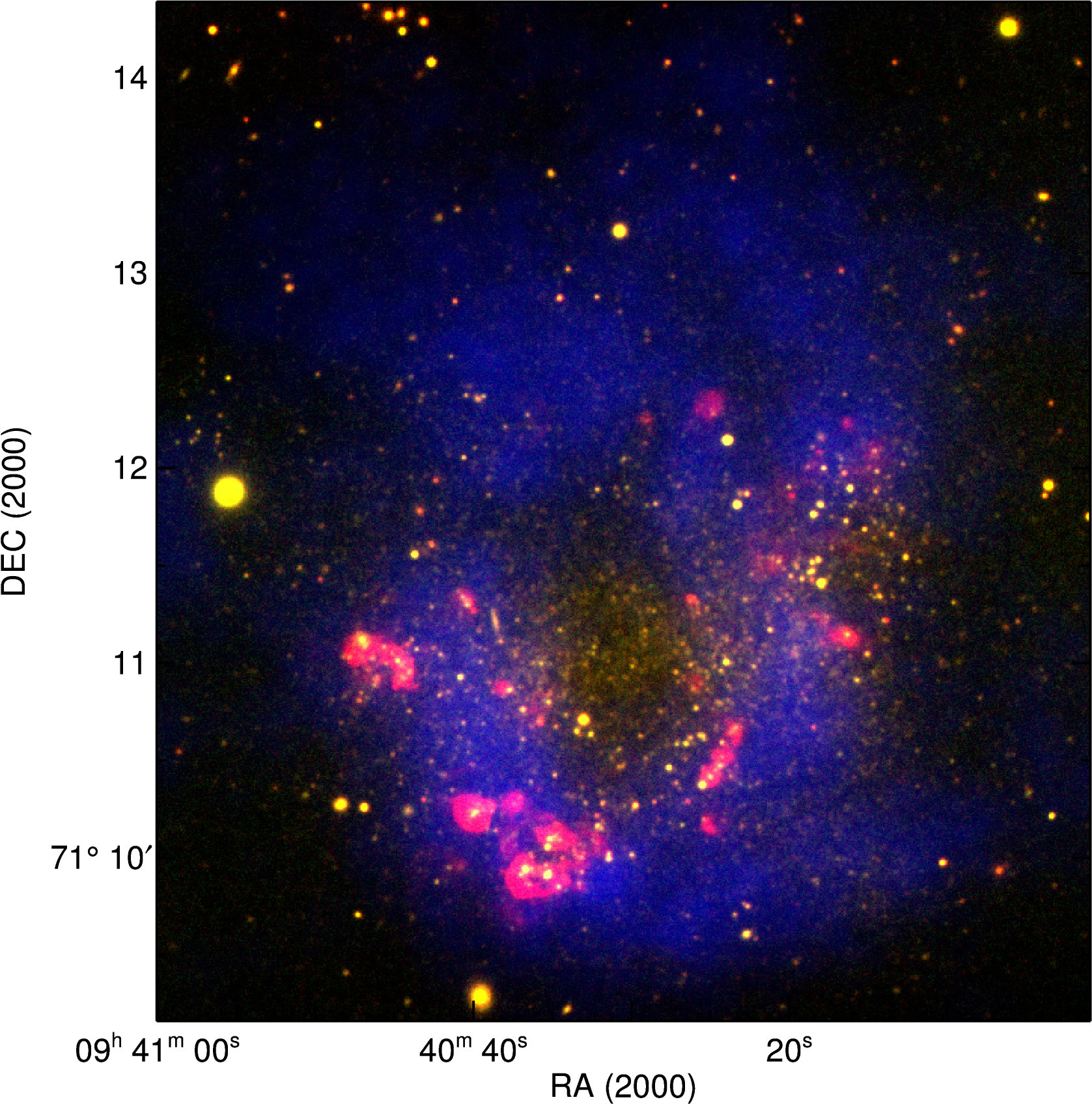}~\includegraphics[width=0.5\linewidth]{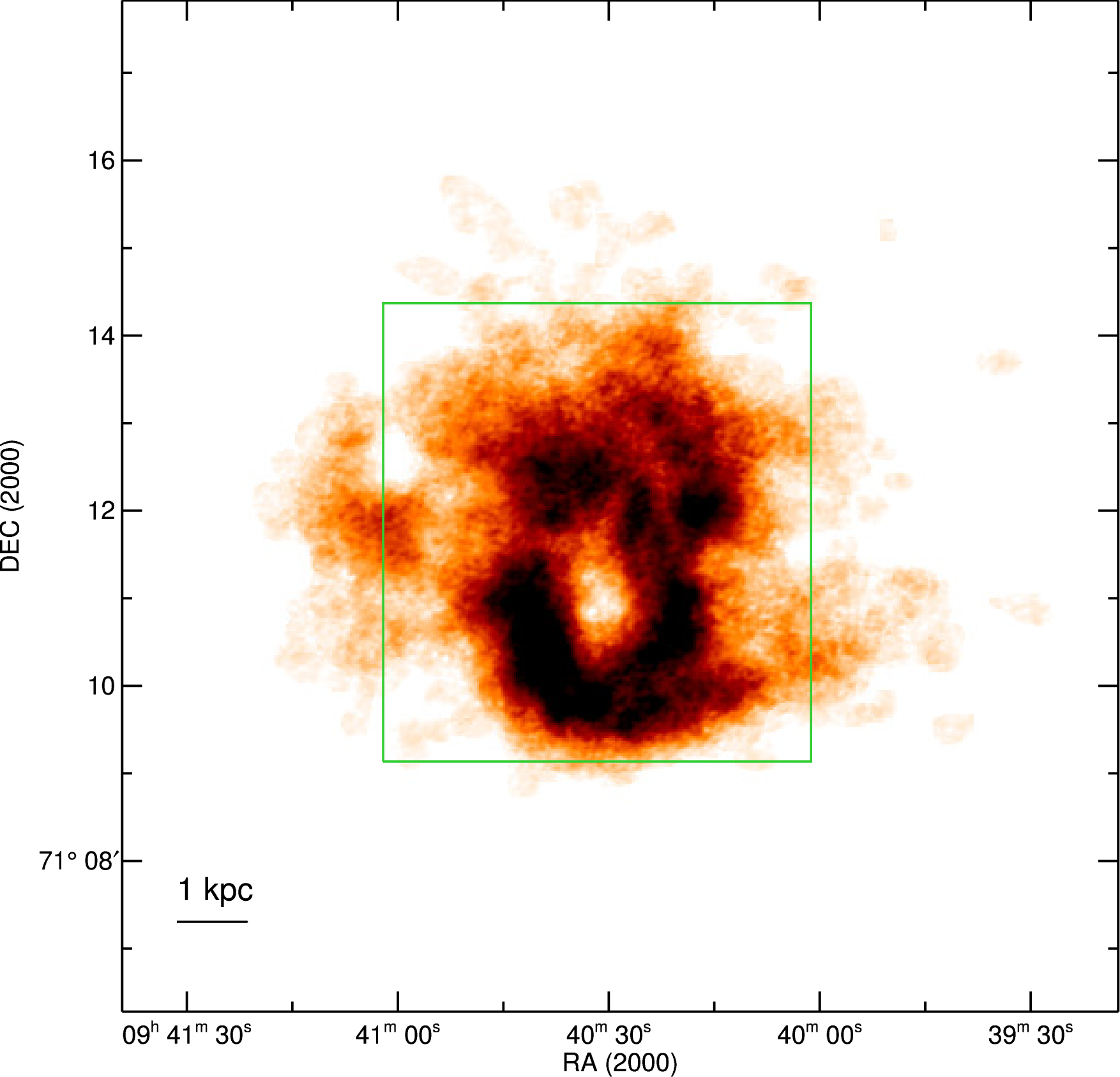}
\caption{Holmberg I. Left: False-color composite image of distribution of H$\alpha$ (red color) and stellar continuum (yellow color) obtained with the SAO RAS 6-m telescope, and HI 21 cm emission (LITTLE THINGS, blue color). Right: HI distribution obtained by LITTLE THINGS data. Green rectangle denotes the region shown in the left panel.
}\label{fig:hoi}
\end{figure}

In this paper we have discussed  the nature of triggered star formation complexes located in the rims of the giant (1 kpc and more) HI supershells. These structures are common in nearby irregular galaxies. Here we considered three galaxies: IC~1613, IC~2574 and Holmberg~II, that might be best examples of the triggering of star formation in supergiant shells at the different evolution stages. Gas kinematics analysis  gave us evidence about the influence of ongoing star formation onto HI supershells. We observe weak and faint filaments of ionized gas outside HII regions that might be a consequence of the ionizing photons leakage from embedded star formation regions. This process might perturb HI shells and lead to their destruction at the later stage. In the final consequence of that process, the ISM of the irregular galaxy might represent a single neutral supershell with diameter about the disc size and the ongoing star formation in its rim. A picture resembling this is observed in Holmberg~I galaxy (see Fig.~\ref{fig:hoi}).

Large numbers of HI supershells in galaxy discs is a common picture for nearby irregular galaxies, but the star formation is not distributed uniformly inside them. It seems that the collision of the HI supershells might be one of the main drivers of the star formation triggering on such large scales.

%The H$\alpha$ emission often observed inside the small HI shells, but it was never  before observed inside kpc-sized HI %supershells where young stellar population is absent.
We found for the first time diffuse ionized gas inside the SGS in IC~2574 that shows the signs of the expansion and have size similar to diameter of the SGS. Similar and more confidently detected structures were discovered in Holmberg~II galaxy. The structure of the H$\alpha$ supershell in that galaxy also coincides with the inner wall of the largest HI supershell with active ongoing star formation inside. We considered two mechanisms of the formation of such structures: ionization of the inner wall of neutral supershell by the stellar population inside and/or by the leakage of the ionizing photons from the bright HII complexes.

\section*{Acknowledgements}
This work was supported by the Russian Foundation for Basic Research (project 14-02-00027) and by a grant from the President of the Russian Federation (MD3623.2015.2). A. Moiseev is grateful for the financial support of the Dynasty Foundation.
The observations at the 6-meter BTA telescope were carried out with the financial support of the Ministry of Education and Science of the Russian Federation
(agreement No. 14.619.21.0004, project ID RFMEFI61914X0004).

\end{document}